\documentclass[preprint,12pt]{elsarticle}
\usepackage{graphicx}
\usepackage{enumerate}
\usepackage{amsmath}
 \usepackage{framed}
  \usepackage{setspace}
  \usepackage{amssymb}
  \usepackage[justification=centering]{caption}
\usepackage{algorithm}
\usepackage{algpseudocode}
\usepackage{comment}
\usepackage{bm}

\newtheorem{theorem}{Theorem} 
\newtheorem{definition}{Definition}
\newtheorem{lemma}{Lemma}

\makeatletter
\renewcommand{\ALG@beginalgorithmic}{\normalsize}
\makeatother

\usepackage{float}
\usepackage{lipsum}

\makeatletter
\newenvironment{breakablealgorithm}
  {
   \begin{center}
     \refstepcounter{algorithm}
     \hrule height.8pt depth0pt \kern2pt
     \renewcommand{\caption}[2][\relax]{
       {\raggedright\textbf{\ALG@name~\thealgorithm} ##2\par}%
       \ifx\relax##1\relax 
         \addcontentsline{loa}{algorithm}{\protect\numberline{\thealgorithm}##2}%
       \else 
         \addcontentsline{loa}{algorithm}{\protect\numberline{\thealgorithm}##1}%
       \fi
       \kern2pt\hrule\kern2pt
     }
  }{
     \kern2pt\hrule\relax
   \end{center}
  }
\makeatother

\begin{document}

\begin{frontmatter}
\title{Quantum impossible differential and truncated differential cryptanalysis}


\author{Huiqin Xie$^{1,2,3}$}
\author{Li Yang$^{1,2,3}$\corref{1}}
\cortext[1]{Corresponding author email: yangli@iie.ac.cn}
\address{1.State Key Laboratory of Information Security, Institute of Information Engineering, Chinese Academy of Sciences, Beijing 100093, China\\
2.Data Assurance and Communication Security Research Center,Chinese Academy of Sciences, Beijing {\rm 100093}, China\\
3.School of Cyber Security, University of Chinese Academy of Sciences, Beijing {\rm  100049}, China}

\begin{abstract}
Traditional cryptography is suffering a huge threat from the development of quantum computing. While many currently used public-key cryptosystems would be broken by Shor's algorithm, the effect of quantum computing on symmetric ones is still unclear. The security of symmetric ciphers relies heavily on the development of cryptanalytic tools. Thus, in order to accurately evaluate the security of symmetric primitives in the post-quantum world, it is significant to improve classical cryptanalytic methods using quantum algorithms. In this paper, we focus on two variants of differential cryptanalysis: truncated differential cryptanalysis and impossible differential cryptanalysis. Based on the fact that Bernstein-Vazirani algorithm can be used to find the linear structures of Boolean functions, we propose two quantum algorithms that can be used to find high-probability truncated differentials and impossible differentials of block ciphers, respectively. We rigorously prove the validity of the algorithms and analyze their complexity. Our algorithms treat all rounds of the reduced cipher as a whole and only concerns the input and output differences at its both ends, instead of specific differential characteristics. Therefore, to a certain extent, they alleviate the weakness of conventional differential cryptanalysis, namely the difficulties in finding differential characteristics as the number of rounds increases.

\end{abstract}


\end{frontmatter}


The development of quantum computing has greatly impacted classical cryptography. Owing to Shor's algorithm \cite{Sho94}, many currently used public-key cryptosystems are insecure against the adversaries in possession of quantum computers, such as RSA, ElGamal and any other cryptosystems based on discrete logarithms or factorization. This has led to the advent of post-quantum cryptography, which studies classical systems that resist quantum adversaries.

In light of the fact that public-key cryptography is suffering from a huge threat due to quantum algorithms, it is natural to consider the impact of quantum attacks on symmetric cryptosystems. A representative example is Grover's search algorithm \cite{Gro96}, which can provide a quadratic speedup for any generic exhaustive key search. This indicates that in the post-quantum world, the key lengths of symmetric primitives need to be doubled to maintain an equivalent ideal security. However, exhaustive key-search attack only defines the ideal security. To understand the real security of symmetric primitives, it is necessary to study how other attacks can be performed by quantum adversaries in the real world. This direction has draw more and more attention in recent years.

Simon's algorithm was proposed in 1997 \cite{Sim97} and has been applied to the analysis of symmetric cryptography. Given a Boolean function with certain promise, one can apply Simon's algorithm to find its periods. Based on this, Kuwakado and Morii constructed a quantum distinguisher for the three-round Feistel scheme \cite{KM10}, which has been proved to be a secure pseudo-random permutation in classical setting \cite{MC88}. Afterwards, they further used Simon's algorithm to extract the key of the Even-Mansour scheme \cite{KM12}. These two results embody the advantages of quantum algorithms in symmetric cryptanalysis. Santoli and Schaffner subsequently extended the result in \cite{KM10} and relaxed the assumption that the internal function of the three-round Feistel scheme must be permutations \cite{SS17}. In the same paper, they also presented a quantum forgery attack to the CBC-MAC scheme. Simultaneously, Kaplan \textit{et al}. also proved that the attack on the Feistel scheme in \cite{KM10} can be extended to the situation where the internal function is not a permutation by a different approach \cite{KLLNP16}. Furthermore, they also used Simon's algorithm to attack other symmetric schemes, such as GMAC, CLOC and so on. These attacks are all in the context of quantum chosen-plaintext attack \cite{BZ13,GHS16}. Roetteler and Steinwandt, however, apply Simon's algorithm to related-key attack \cite{RS15}. They showed that, under certain conditions, the access to query superpositions of related keys to a block cipher enables the attacker to extract the secret key efficiently.

Although these results are striking, they are still not enough to evaluating the actual security of ciphers in post-quantum world. In symmetric cryptography, designers demonstrate the security of schemes by proving that they can resist some specific attacks. The security of ciphers relies heavily on the development of cryptanalytic tools. Therefore, using quantum techniques to improve the main classical analytic tools, such as differential cryptanalysis and linear cryptanalysis, is significant for the design of quantum-secure symmetric cryptosystems.

The idea of applying quantum algorithms in differential cryptanalysis was first considered in \cite{ZLZS15}. The authors applied Grover's algorithm in the key-recovery phase and obtained a quadratic speedup. Afterwards, Kaplan \textit{et al}. further studied the application of Grover's algorithm in cryptanalysis and applied it in the key-recovery phases of various variants of differential and linear attacks \cite{KLLNP17}. On the other hand, Li and Yang focused on the first phase of differential cryptanalysis and applied Bernstein-Vazirani (BV) algorithm \cite{BV97} to finding high-probability differentials of block ciphers \cite{LY15}. Their algorithm has a obvious flaw, which has been solved by Xie and Yang in \cite{XY17}. In this paper, we will further develop these works and apply BV algorithm to truncated differential cryptanalysis and impossible differential cryptanalysis.
\vskip 0.2cm

\noindent
\textbf{Quantum Attackers.} There are two types of quantum attackers that have been studied in previous papers. The first type can perform quantum operations and make classical queries on the cryptographic primitives, denoted as $Q_1$. The second type, in addition to quantum operations and classical queries, can also make quantum queries on the cryptographic oracle, denoted as $Q_2$. That is, $Q_2$-type quantum attackers can query the cryptographic oracle directly with superposition state, and obtains the superposition of the corresponding outputs. The second type is more demanding on the attackers' power, because the access to the quantum oracle of cryptographic primitives is difficult to achieve for the attacker in practice.
\vskip 0.2cm

\noindent
\textbf{Complexity.} In general, the efficiency of an attack can be defined by three parameters: data complexity, time complexity and memory complexity. The data complexity is the amount of queries made by the attacker; the time complexity is the time it takes to execute the attack; and the memory complexity is the memory required for the attack. We assume that a query needs one unit of time, then the data complexity is contained in the time complexity. Therefore, when analyze the complexity of an algorithm, we only need to consider its time complexity and memory complexity. Specifically, time complexity can be divided into two parts: the time required to execute the quantum computing part and the time required to execute the classical computing part. For the first part, since any quantum circuit can be expressed in terms of gates in some reasonable, universal, finite set of unitary quantum gates \cite{NC00}, the corresponding complexity is the number of universal gates. For the second part, the time complexity is equal to the number of elementary operations plus the number of classical queries. (We do not take quantum query into account since all attack algorithms in this paper only need $Q_1$-type attackers.) As for the memory complexity, since the cost of storing quantum information is much greater than classical information, it is expressed as the amount of qubits needed to perform the algorithm.

Based on the above discussion, in this paper we analyze the complexity of a quantum algorithm from three perspectives: the number of universal gates, the time complexity corresponding to the classical computing part and the amount of qubits needed in the algorithm.
\vskip 0.2cm

\noindent
\textbf{Our contributions.} In this paper, we proceed with previous works and further explore how to apply quantum algorithms to classical cryptanalytic tools. We focus on two variants of differential cryptanalysis: truncated differential cryptanalysis and impossible differential cryptanalysis. Based on the fact that BV algorithm can be used to find the linear structures of Boolean functions, we propose two quantum algorithms for finding high-probability truncated differentials and impossible differentials. Afterwards, we rigorously prove the validity of the algorithms and analyze its complexity. The amounts of universal gates and qubits needed by these two algorithms are both polynomials of $n$, where $n$ is the blocksize.

The proposed algorithms for finding high-probability truncated differentials and impossible differentials have following advantages:
\vskip 0.2cm

\noindent
$\bullet\,$ In conventional truncated differential cryptanalysis, the attacker finds high-probability truncated differentials by searching for a high-probability truncated differential characteristic. However, as the number of rounds increases, the probability of differential characteristics will usually decreases dramatically. Therefore, using traditional method to find high-probability truncated differentials will become more and more difficult as the number of rounds increases. Classical miss-in-the middle technique has the similar problem, because it finds probability-1 differentials also based on differential characteristics. By contrast, our algorithms only concerns the input and output differences at both ends of the cipher, instead of a specific differential characteristic, so the increase in the number of rounds has a much smaller effect on them. Therefore, our algorithms should work better than traditional methods when the number of rounds of the block cipher is large.
\vskip 0.2cm

\noindent
$\bullet\,$ Since traditional methods of finding high-probability truncated differentials is based on differential characteristics, it can only find the high-probability truncated differentials whose input and output differences can be connected by a high-probability differential characteristic. However, a high-probability truncated differential does not imply a high-probability truncated differential characteristic. This kind of truncated differentials is actually restrictive. By contrast, our algorithm only concerns the input and output differences at both ends, and thus can find more general high-probability truncated differentials.
\vskip 0.2cm

\noindent
$\bullet\,$ The proposed quantum algorithms do not require quantum queries. This makes the differential attacks based on these algorithms more practical. Being able to query the cryptographic oracle in quantum superpositions is actually a strong requirement for the attacker's ability. Many recently proposed quantum algorithms for attacking block ciphers, such as the algorithms in \cite{KLLNP16,KM10,KM12,SS17}, require quantum queries. Compared with these algorithms, our algorithms are easier to implement for quantum attackers in practice.
\vskip 0.2cm

\noindent
\textbf{Related work.} In \cite{KLLNP17}, Kaplan \textit{et al}. also studied the quantum truncated differential cryptanalysis. They applied Grover's algorithm in the key recovering phase, while we focus on the first phase of truncated differential cryptanalysis, namely, finding a high-probability truncated differential. Impossible differential attack is investigated both in \cite{XY17} and in this paper. The quantum algorithm proposed in \cite{XY17}, however, finds impossible differentials directly, and can only find the impossible differentials that have a special structure, while our algorithm uses the miss-in-the-middle technique and can find more general impossible differentials.

\section{Preliminaries}
Throughout this paper, we denote an arbitrary block cipher as $E$, and assume it can be implemented efficiently as a quantum circuit. Since any quantum circuit can be expressed in terms of gates in some universal, finite set of unitary quantum gates \cite{NC00}, $E$ can be executed by a quantum circuit composed of gates in the universal set. Let $|E|_Q$ denote the amount of universal gates in the circuit, and $n$ denote the blocksize of $E$. If $E$ can be computed in polynomial time with respect to $n$, then $|E|_Q$ is a polynomial of $n$.
\subsection{Linear structure of Boolean functions}
Let $\mathbb{F}_2=\{0,1\}$ denote a finite field of characteristic 2. $\mathbb{F}_2^n$ is the $n$-dimension vector space over $\mathbb{F}_2$. $\mathcal{B}_n$ denotes the set of all Boolean functions from $\mathbb{F}_2^n$ to $\mathbb{F}_2$. The notion of linear structure has been studied for their cryptanalytic significance \cite{Dub01,LY18,OK94}, and is defined as follows:
\begin{definition}
Suppose $f$ is a Boolean function in $\mathcal{B}_n$. A vector $a\in \mathbb{F}_2^n$ is called a linear structure of $f$ if
$$
f(x\oplus a)+f(x)=f(a)+f(\bm{0}),\,\,\,\forall x\in \mathbb{F}_2^n,
$$
where $\oplus$ denotes the bitwise exclusive-or, and $+$ denotes the addition operation in $\mathbb{F}_2$.
\end{definition}

Let $U_f$ denote the set of all linear structures of $f$, and
$$
U_f^i:=\{a\in \mathbb{F}_2^n|f(x\oplus a)+f(x)=i,\, \forall x\in \mathbb{F}_2^n\}
$$
for $i=0,1$. It is obvious that $U_f=U_f^0\cup U_f^1$. The linear structure of a Boolean function has a close relation with its Walsh spectrum, whose definition is:
\begin{definition}Suppose $f\in\mathcal{B}_n$, the Walsh spectrum of $f$ is also a function in $\mathcal{B}_n$, and is defined as
\begin{align*}
S_f:\mathbb{F}_2^n&\longrightarrow\mathbb{F}_2\\
\omega&\longrightarrow S_f(\omega)=\frac{1}{2^n}\sum_{x\in \mathbb{F}_2^n}(-1)^{f(x)+\omega\cdot x}.
\end{align*}
\end{definition}

For any function $f\in\mathcal{B}_n$, let $N_f=\{\omega\in \mathbb{F}_2^n|S_f(\omega)\neq0\}$. The following lemma shows the link between the Walsh spectral and the linear structure.
\begin{lemma}[\cite{Dub01}, Corollary 1]For any $f\in\mathcal{B}_n$, $i\in\{0,1\}$, it holds that
$$
U_f^i=\{a\in \mathbb{F}_2^n|\,\omega\cdot a=i,\forall\omega\in N_f\}.
$$
\end{lemma}

According to Lemma 1, given a sufficiently large subset $H$ of $N_f$, one may be able to obtain $U_f^i$ by solving the linear system $\{x\cdot\omega=i|\omega\in H\}$. (Solving the linear system $\{x\cdot\omega=i|\omega\in H\}$ means finding vectors $x$ such that $x\cdot\omega=i$ holds for $\forall\omega\in H$.)

For any positive integers $m,n$, let $\mathcal{C}_{m,n}$ denote the set of Boolean functions from $\mathbb{F}_2^m$ to $\mathbb{F}_2^n$. The linear structure of the functions in $\mathcal{C}_{m,n}$ is defined as following:
\begin{definition}
Suppose $F\in\mathcal{C}_{m,n}$. A vector $a\in \mathbb{F}_2^m$ is said to be a linear structure of $F$ if there is a vector $\alpha\in\mathbb{F}_2^n$ such that
$$
F(x\oplus a)\oplus F(x)=\alpha,\,\,\,\forall x\in\{0,1\}^m.
$$
\end{definition}

Suppose $F=(F_1,F_2,\cdots,F_n)$. It is obvious that a vector is a linear structure of $F$ if and only if it is a linear structure of $F_j$ for all $j=1,2,\cdots,n$. Thus, we can obtain the linear structures of $F$ by first finding the linear structures of every component function $F_j$ separately, and then picking out the common ones. Let $U_F$ be the set of the linear structures of $F$, and $U_F^{\alpha}=\{a\in \mathbb{F}_2^m|F(x\oplus a)\oplus F(x)=\alpha,\,\forall x\}$. It is obvious that $U_F=\cup_{\alpha}U_F^{\alpha}$.

\subsection{Bernstein-Vazirani algorithm}

BV algorithm was proposed by Bernstein and Vazirani \cite{BV97}. It solves the following problem: given the permission to query a function $f(x)=a\cdot x$ with superposition states, where $a\in\{0,1\}^n$ is a secret string, find $a$. BV algorithm works as follows:
\begin{enumerate}[  1.]
\item Apply the Hadamard operator $H^{(n+1)}$ to the initial state $|\psi_0\rangle=|0\rangle^{\otimes n}|1\rangle$ to obtain the state $|\psi_1\rangle=\sum_{x\in \mathbb{F}_2^n}\frac{|x\rangle}{\sqrt{2^n}}\cdot\frac{|0\rangle-|1\rangle}{\sqrt{2}}$.

\item Query the quantum oracle of $f$ , giving the state $|\psi_2\rangle=\sum_{x\in \mathbb{F}_2^n}\frac{(-1)^{f(x)}|x\rangle}{\sqrt{2^n}}\frac{|0\rangle-|1\rangle}{\sqrt{2}}$.
\item Discard the last qubit, and apply the Hadamard gates $H^{(n)}$ to the rest $n$ qubits again to obtain the state
\begin{align}
|\psi_3\rangle=\sum_{y\in \mathbb{F}_2^n}(\frac{1}{2^n}\sum_{x\in \mathbb{F}_2^n}(-1)^{f(x)+y\cdot x})|y\rangle.
\end{align}
Since $f(x)=a\cdot x$, we have
\begin{align*}
|\psi_3\rangle&=\sum_{y\in \mathbb{F}_2^n}(\frac{1}{2^n}\sum_{x\in \mathbb{F}_2^n}(-1)^{(a\oplus y)\cdot x})|y\rangle=|a\rangle.
\end{align*}
Thus, measuring $|\psi_3\rangle$ in the computational basis gives the vector $a$ with a probability of one.
\end{enumerate}

According to Eq.(1), if we apply the BV algorithm to a general function $f$ in $\mathcal{B}_n$, the resulting state before measurement will be $\sum_{y\in \mathbb{F}_2^n}S_f(y)|y\rangle$, where $S_f(\cdot)$ is the Walsh spectrum of $f$. By measuring this state in the computational basis, we will get a vector $y\in\mathbb{F}_2^n$ with probability $S_f(y)^2$. Therefore, running BV algorithm on $f$ always gives a vector in $N_f$.

Let $|f|_Q$ be the number of universal gates in the quantum circuit that implements $f$. Running BV algorithm on $f$ requires a total of $2n+1+|f|_Q$ universal gates and one quantum query. The amount of qubits needed to perform BV algorithm is $n+1$. Based on Lemma 1 and the fact that applying BV algorithm to $f$ always gives a vector in $N_f$, a quantum algorithm that finds nonzero linear structures was proposed in \cite{LY18}:

\begin{algorithm}[H]
\caption{}
\begin{algorithmic}[1]
\Require Let $p(n)$ be an arbitrary polynomial of $n$. $\Phi$ denotes the null set. Initialize the set $H:=\Phi$.
 \For{$p=1,2,\cdots,p(n)$}

   \State run BV algorithm on $f$ to obtain an $n$-bit output $\omega\in N_f$

       \State let $H=H\cup \{\omega\}$;
\EndFor
\State  solve the linear systems $\{x\cdot \omega=i|\omega\in H\}$ to obtain solutions $A^i$ for $i=0,1$, respectively;
\If{ $A^0\cup A^1\subseteq\{\bm{0}\}$}  \State output ``No'' and halt;
 \Else \State output $A^0$ and $A^1$;
 \EndIf
 \end{algorithmic}
\end{algorithm}

For an arbitrary function $f\in\mathcal{B}_n$, let
\begin{equation}
\delta'_f=\frac{1}{2^n} \max_{\substack{a\in \mathbb{F}_2^n\\ a\notin U_f}}\max_{i\in \mathbb{F}_2}|\{x\in \mathbb{F}_2^n|f(x\oplus a)+f(x)=i\}|.
\end{equation}
It is obvious that $\delta'_f<1$. Intuitively, the smaller $\delta'_f$ is, the better it is to rule out the vectors that are not the linear structure of $f$ when running Algorithm 1. The following two theorems justify the validity of Algorithm 1.
\begin{theorem}[\cite{LY18}, Theorem 4.1] If applying Algorithm 1 to a function $f\in\mathcal{B}_n$ outputs sets $A^0$ and $A^1$, then for any vector $a\in A^i$ ($i=0,1$), any $\epsilon$ satisfying $0<\epsilon<1$, we have
\begin{equation}
{\rm Pr}\Big[1-\frac{|\{x\in \mathbb{F}_2^n|f(x\oplus a)+f(x)=i\}|}{2^n}<\epsilon\Big]>1-e^{-2p(n)\epsilon^2}.
\end{equation}
\end{theorem}

\begin{theorem}[\cite{XY17}, Theorem 2] Suppose $f\in\mathcal{B}_n$ and $\delta'_f\leq p_0<1$ for some constant $p_0$. If applying Algorithm 1 to $f$ with $p(n)=n$ queries returns the sets $A^0$ and $A^1$, then for any $a\notin U_f^i$ ($i=0,1$), we have that
$$
{\rm Pr}[a\in A^i]\leq p_0^{n}
$$
\end{theorem}

These two theorems are proved in \cite{LY18} and \cite{XY17}, respectively. Theorem 1 states that the vectors in $A^0$ and $A^1$ have a high probability of being approximate linear structures of $f$. Theorem 2 demonstrates that, under assumption that $\delta'_f\leq p_0<1$, the probability of Algorithm 1 after $O(n)$ quantum queries outputting a vector that is not a linear structure of $f$ is negligible.

Although Algorithm 1 needs to query the quantum oracle of $f$, we stress that, the quantum algorithms for finding high-probability truncated differentials and impossible differentials proposed in this paper do not require quantum queries.

\subsection{Differential cryptanalysis}
Differential cryptanalysis was proposed by Biham and Shamir \cite{BS91}, which exploits the existence of high-probability differentials. Let $E$ denote a $r$-round block cipher and $n$ be the blocksize of $E$. For $t\in\{1,\cdots,r\}$, $E^{(t)}$ denotes the reduced version of $E$ with $t$ rounds. Let $F$ denote the function that maps the plaintext to the input of the last round of $E$, i.e. $F=E^{(r-1)}$, and $\mathcal{K}$ be the key space of $F$. The input of $F$ includes a plaintext block and a key in $\mathcal{K}$. When fix a specific key $k\in\mathcal{K}$, the action of $F$ on a block $x$ is denoted by $F_{k}(x)$. Suppose $F_k(x)=y$, $F_k(x')=y'$, then the input difference is given by $\Delta x=x\oplus x'$ and the output difference is given by $\Delta y=y\oplus y'$. The pair $(\Delta x,\Delta y)$ is called a differential of $F_k$. Differential cryptanalysis can be divided into two stages: (I) finding some high-probability differential of $F_k$; and (II) using the found high-probability differential to recover the subkey of the last round.

Several variants of differential cryptanalysis have been developed, such as truncated differential attack \cite{Knu94} and impossible differential attack \cite{BBS99}. They all exploit some  non-random statistical patterns in the distribution of cipher difference. According to the definition of linear structure, if $F_k$ has a linear structure $x\in U_{F_k}^{\alpha}$ for some $\alpha\in\mathbb{F}_2^n$, then $(x,\alpha)$ is also a probability-1 differential of $F_k$. Likewise, if $x$ is an approximate linear structure of $F_k$, then there is a vector $\alpha$ such that $(x,\alpha)$ is a high-probability differential of $F_k$. In light of this relation between linear structures and differentials, we can apply Bernstrin-Vazirani algorithm in various variants of differential cryptanalysis.

\section{Quantum truncated differential cryptanalysis}

Truncated differential cryptanalysis was introduced by Knudsen \cite{Knu94} and has been applied to many symmetric cryptosystems \cite{KB96,KRW99,Knu94}. In a conventional differential attack, the attacker considers the full difference of two texts, while the truncated differential cryptanalysis analyzes differences that are only partially determined. That is, the attacker only predicts part of the bits instead of the entire block.

We still consider a $r$-round block cipher $E$ and its reduced version $F=E^{(r-1)}$. Let $\mathcal{K}$ and $\mathcal{S}$ be the key spaces of the first $r-1$ rounds and the last round of $E$, respectively. When fix a specific key $k\in\mathcal{K}$, the action of $F$ on a block $x$ is denoted by $F_{k}(x)$. Suppose $(\Delta x,\Delta y)$ is a differential of $F_k$. If $\Delta x'$ is a subsequence of $\Delta x$ and $\Delta y'$ is a subsequence of $\Delta y$, then $(\Delta x',\Delta y')$ is called a truncated differential of $F_k$. In this paper, we only consider the case in which the input difference is a full difference, i.e. $\Delta x'=\Delta x$. The bits that appear in $\Delta y'$ are called predicted bits, and the others are called unpredicted bits. If a full output difference $\Delta z$ satisfies that the predicted bits of $\Delta y'$ are equal to the corresponding bits of $\Delta z$, then we say $\Delta z$ matches $\Delta y'$, denoted $\Delta z=\Delta y'$.

Truncated differential cryptanalysis consists of two stages. In the first stage, the attacker searches for a high-probability truncated differential $(\Delta x,\Delta y')$ of $F_k$. In the second stage, the attacker recovers the key of the last round using the found truncated differential $(\Delta x,\Delta y')$. Specifically, he fixes the input difference $\Delta x$ and makes classical queries to obtain $2N$ ciphers. Then for each $s\in\mathcal{S}$, he decrypts the last round to get $N$ output differences of $F_k$, and counts the number of them that match $\Delta y'$. The correct key is likely to be the one with the maximum count.


The success probability and the number of ciphertext pairs needed in the counting scheme are related to the signal to noise ratio \cite{BS91}, which is defined as:
$$
S/N=\frac{|\mathcal{S}|\times p}{\gamma\times \lambda},
$$
where $|\mathcal{S}|$ is the number of possible candidate keys of the last round, $p$ is the probability of the used truncated differential, $\gamma$ is the average count contributed by each pair of plaintexts, and $\lambda$ is the ratio of non-discarded pairs to all pairs. We only consider the case in which $\lambda=1$. If $S/N\leq1$, then the truncated differential attack will not succeed. Thus, we need to find truncated differentials whose the signal to noise ratio is larger than 1. The larger $S/N$ is, the better it is for recovering the key. Further details about $S/N$ can be found in Appendix A.

The quantum algorithm we will propose is applied to the first stage of truncated differential cryptanalysis, i.e. finding a high-probability truncated differential. In traditional truncated differential cryptanalysis, since the key of the reduced cipher $F_k$ is unknown to the attacker, he actually needs to find a high-probability truncated differential that is independent of the key, namely, a truncated differential that has a high probability regardless of the value of $k$. Even though our quantum algorithm cannot find key-independent high-probability truncated differentials, it can find truncated differentials which have high probability for most of the keys in $\mathcal{K}$. In more detail, for any polynomial $q(n)$, the attacker can execute the algorithm properly so that the output truncated differentials have high probability for more than $(1-\frac{1}{q(n)})$ proportion of the keys in $\mathcal{K}$.

We present the algorithm in Section 3.1, and analyze its validity, complexity and advantages in Section 3.2.

\subsection{Quantum algorithm for finding truncated differentials}

Suppose the reduced cipher $F_k=(F_{k,1},F_{k,2},\cdots,F_{k,n})$. Intuitively, we can use Algorithm 1 to find the approximate linear structures of each component function $F_{k,j}$ ($1\leq j\leq n$). Every approximate linear structure of $F_{k,j}$ can induce a high-probability differential of it. If there exist several component functions having the same approximate linear structure, then it can induce a high-probability truncated differential of $F_k$. The positions of the predicted bits correspond to the positions of these component functions. However, the problem is that calling Algorithm 1 requires to query $F_k$ with quantum superpositions. This cannot be done since the attacker does not know the value of $k$. In traditional differential attack, the attacker cannot query the reduced cipher $F_k$, either. He therefore analyze the properties of the block cipher and tries to find high-probability truncated differentials independent of the key. This inspires us to also look for key-independent high-probability truncated differentials, or at least, the truncated differentials whose probability is high for most of the keys.

To achieve this goal, we consider the reduced cipher $F$ without specifying the key. Suppose $\mathcal{K}=\{0,1\}^m$. The function
\begin{align*}
F:\{0,1\}^{n}\times\{0,1\}^m&\longrightarrow \{0,1\}^n\\
(\,\,\,x\,\,\,\,\,\,,\,\,\,\, k\,\,\,)\quad&\longrightarrow F_{k}(x)
\end{align*}
is determined and known to the attacker. Therefore, the attacker can use a quantum circuit to implement the operator $$U_F:|x,k\rangle|y\rangle\rightarrow|x,k\rangle|y\oplus F(x,k)\rangle$$ efficiently. The number of universal gates in the circuit is denoted by $|F|_Q$, which is a polynomial of $n$ and $m$. Let $F=(F_1,\cdots,F_n)$. Each component function $F_j$ can also be implemented by efficient quantum circuit, and the number of universal gates of the corresponding circuit is denoted by $|F_j|_Q$. It is obvious that $\sum_{j=1}^n|F_j|_Q=|F|_Q$. Since the attacker can perform the quantum circuit of each $F_j$ by himself, he can execute Algorithm 1 on $F_j$ without the need for quantum queries. Therefore, the attacker can use Algorithm 1 to find high-probability truncated differentials of $F$. He first runs Algorithm 1 to obtain the approximate linear structures of each $F_j$, then chooses a vector that is common approximate linear structure of multiple component functions $F_j's$ as the input difference of the truncated differential.

However, our goal is not to find a high-probability truncated differential of $F$, but to find a high-probability truncated differential of $F_k$. Note that the input of $F_j$ includes a plaintext block and a key in $\mathcal{K}$, so the approximate linear structure we obtain by applying Algorithm 1 on $F_j$ has $n+m$ bits, including $n$-bit plaintext difference $\Delta x$ and $m$-bit key difference $\Delta k$. In order to use it to induce a differential of $F_k$, the key difference $\Delta k$ needs to be zero. To this end, we only need to discard the last $m$ bits of the output vector when calling BV algorithm. Specifically, the quantum algorithm for finding high-probability truncated differentials of $F_k$ is as follows:

\begin{breakablealgorithm}[H]
\caption{}
\begin{algorithmic}[1]
\Require The quantum circuit for implementing each $F_j$ $(1\leq j\leq n)$ is given. $q(n)$ is an arbitrary polynomial chosen by the attacker. $\sigma\in[0,1)$ is a constant and is also chosen by the attacker. Let $p(n)=\frac{1}{2\sigma^2}q(n)^2n^3$ and initialize the set $H:=\Phi$.
\vspace{0.2cm}
\For{$j=1,2,\cdots,n$}
 \For{$p=1,\cdots,p(n)$}
  \State run BV algorithm on $F_j$ to obtain an $(n+m)$-bit output $\omega=$
  \State$(\omega_1,\cdots,\omega_n,\omega_{n+1},\cdots,\omega_{n+m})\in N_{F_j}$;
  \State let $H=H\cup\{(\omega_1,\cdots,\omega_n)\}$;
 \EndFor
 \State solve the linear systems $\{x\cdot \omega=i_j|\omega\in H\}$ to obtain the set $A_j^{i_j}$ for $i_j=0,1$,
 \State respectively;
 \State Let $A_j= A_j^0\cup A_j^1$ and $H=\Phi$;
\EndFor
\For{$t=n,n-1,\cdots,1$}
 \If{$S/N=2^t(1-\sigma)>1$}
 \If{there exist $t$ different $j_1,\cdots,j_t$ s.t. $A_{j_1}\cap\cdots\cap A_{j_t}\supsetneq\{\bm{0}\}$}
 \State choose an arbitrary nonzero vector $a\in A_{j_1}\cap\cdots\cap A_{j_t}$;
 \State Let
 $$
b_{j}=\left\{\begin{array}{cc}
i_j, & \,\,j\in\{j_1,\cdots,j_t\}\\
\times, &  \,\,j\notin\{j_1,\cdots,j_t\},\\
\end{array}
\right.
$$
 \State where $j=1,2,\cdots,n$, and $i_j$ is the superscript such that $a\in A_j^{i_j}$;
 \State Let $b=(b_1,\cdots,b_n)$, output $(a,b)$ and stop;
 \EndIf
 \Else \State output ``No'' and stop;
 \EndIf
\EndFor
\State Output ``No'' and stop;
 \end{algorithmic}
\end{breakablealgorithm}

Lines 1-10 of Algorithm 2 are for finding the approximate linear structures of $F_j$ for each $j\in\{1,2,\cdots,n\}$. Lines 11-22 are for picking out a vector such that it is a common approximate linear structure of as many $F_j$ as possible. The output $(a,b)$ is a truncated differential of $F_k$. The symbol ``$\times$'' in vector $b$ denotes the unpredicted bits. When the attacker executes a truncated differential attack, he first chooses a constant $\sigma$ and a polynomials $q(n)$, then runs Algorithm 2 to obtain a truncated differential $(a,b)$. According to Theorem 3 presented in the next subsection, $(a,b)$ is a high-probability truncated differential of $F_k$ for most keys in $\mathcal{K}$. In more detail, except for a negligible probability, for more than $(1-\frac{1}{q(n)})$ of the keys in $\mathcal{K}$, the truncated differential $(a,b)$ has a probability greater than $1-\sigma$.

To justify the feasibility of recovering the key of the last round using $(a,b)$, we need to demonstrate that the  signal to noise ratio $S/N$ is greater than 1. To do this, we first estimate the value of $\gamma$, i.e. the average count contributed by each plaintext pair. Suppose $t$ bits of $b$ are predicted, there are $2^{n-t}$ output differences matching $b$ in total. In counting scheme, each pair of plaintexts will be decrypted by $|\mathcal{S}|$ keys separately. The corresponding $|\mathcal{S}|$ output differences obtained by this process can be seen as random. Thus for each pair of plaintexts, there are $$\gamma=\frac{2^{n-t}}{2^n}\times|\mathcal{S}|=\frac{|\mathcal{S}|}{2^t}$$ keys counted on average. Therefore,
$$
N/S\geq\frac{|\mathcal{S}|\times(1-\sigma)}{\frac{|\mathcal{S}|}{2^t}\times1}=2^t(1-\sigma)>1.
$$
The last ``$>$'' holds since we require that $2^t(1-\sigma)>1$ in Algorithm 2 (line 12). After obtaining the truncated differential $(a,b)$, the attacker can use it to recover the subkey of the last round as in classical truncated differential cryptanalysis. This attack works for more than $(1-\frac{1}{q(n)})$ of the keys in $\mathcal{K}$. Even if Algorithm 2 outputs ``No'', the attacker can adjust the values of $q(n)$, $\sigma$ and try again.

\subsection{Analysis of the algorithm }

In this section, we discuss the validity, complexity and advantages of Algorithm 2.

\subsubsection{Validity}
To demonstrate the validity of Algorithm 2, we first give the following theorem:
\begin{theorem}If running Algorithm 2 returns a truncated differential $(a,b)$, then except for a negligible probability, there exists a subset $\mathcal{K}'\subseteq\mathcal{K}$, such that $|\mathcal{K}'|/|\mathcal{K}|>1-\frac{1}{q(n)}$, and for each $k\in\mathcal{K}'$, it holds that
$$
\frac{|\{x\in \mathbb{F}_2^n|F_k(x\oplus a)+F_k(x)=b\}|}{2^n}>1-\sigma.
$$
That is, except for a negligible probability, the output $(a,b)$ is a truncated differential of $F_k$ whose probability is greater than $1-\sigma$ for more than $(1-\frac{1}{q(n)})$ of the keys in $\mathcal{K}$. ( Here,``$=$'' means that $F_k(x\oplus a)+F(x)$ matches $b$. )
\end{theorem}

\noindent
\textbf{Proof}. Suppose $t$ bits of $b$ is predicted, and the corresponding positions of the predicted bits are $j_1,\cdots,j_t$. Let $(a\|0,\cdots,0)$ be the vector obtained by appending $m$ zeros to $a$. Since $a\cdot(\omega_1,\cdots,\omega_n)=0$ implies that $$(a\|0,\cdots,0)\cdot(\omega_1,\cdots,\omega_n,\omega_{n+1},\cdots,\omega_{n+m})=0,$$
the vector $(a\|0,\cdots,0)$ can be viewed as an output of Algorithm 1 on $F_{j_s}$ for all $s=1,2,\cdots,t$. According to Theorem 1, we have that
\begin{equation}
\frac{|\{z\in \mathbb{F}_2^{n+m}|F_{j_s}(\,z\oplus(a\|0,\cdots,0)\,)\oplus F_{j_s}(z)=b_{j_s})\}|}{2^{n+m}}>1-\epsilon,\,\,\forall s=1,2\cdots,t
\end{equation}
holds with a probability greater than $(1-e^{-2p(n)\epsilon^2})^t$. If Eq.$(4)$ holds, we have that the amount of $z$ satisfying
\begin{equation}
F_{j_s}\big(z\oplus(a\|0,\cdots,0)\big)\oplus F_{j_s}(z)=b_{j_s}
\end{equation}
for both $s=1$ and $s=2$ is more than $2^{n+m}[2(1-\epsilon)-1]=2^{n+m}(1-2\epsilon)$. Likewise, the amount of $z$ satisfying Eq.$(5)$ for all $s=1,2,3$ is more than $2^{n+m}(1-3\epsilon)$. By induction, it is easy to verify that the amount of $z$ satisfying Eq.$(5)$ for all $s=1,2,\cdots,t$ is more than $2^{n+m}(1-t\epsilon)$. Thus, with a probability greater than $(1-e^{-2p(n)\epsilon^2})^t$, it holds that
\begin{equation*}
\frac{|\{z\in \mathbb{F}_2^{n+m}|F(\,z\oplus(a\|0,\cdots,0)\,)\oplus F(z)=b)\}|}{2^{n+m}}>1-t\epsilon.
\end{equation*}
Here, ``='' means that $F(\,z\oplus(a\|0,\cdots,0)\,)\oplus F(z)$ matches $b$. Therefore, with a probability greater than $(1-e^{-2p(n)\epsilon^2})^t$, it holds that
\begin{equation}
\frac{|\{(x,k)\in \mathbb{F}_2^n\times \mathbb{F}_2^m|F_k(x\oplus a)\oplus F_k(x)=b\}|}{2^{n+m}}>1-t\epsilon.
\end{equation}
Let
$$
V(k)=\frac{|\{x\in \mathbb{F}_2^n|F_k(x\oplus a)+F_k(x)=b\}|}{2^n}.
$$
Eq.$(6)$ indicates that $\mathbb{E}_k[V(k)]>1-t\epsilon$, where $\mathbb{E}_k[V(k)]$ is the expectation of $V(k)$ when $k$ is chosen randomly and uniformly from $\{0,1\}^m$. Thus, if Eq.$(6)$ holds, then for any polynomial $q(n)$, we have
$$
{\rm Pr}_k\big[\,V(k)>1-q(n)t\epsilon\,\big]>1-\frac{1}{q(n)}.
$$
Otherwise, we have ${\rm Pr}_{k}[1-V(k)\geq q(n)t\epsilon]\geq\frac{1}{q(n)}$, then
\begin{align*}
&\mathbb{E}_k[V(k)]\\
=&1-\mathbb{E}_k[1-V(k)]\\
\leq&1-\frac{1}{q(n)}\cdot q(n)t\epsilon\\
=&1-t\epsilon,
\end{align*}
which contradicts that $\mathbb{E}_k[V(k)]>1-t\epsilon$. Therefore, if Eq.$(6)$ holds, then for more than $(1-\frac{1}{q(n)})$ of the keys in $\mathcal{K}$, we have $V(k)>1-q(n)t\epsilon$. Let $\mathcal{K}'$ be the set of these keys, then $|\mathcal{K}'|/|\mathcal{K}|>1-\frac{1}{q(n)}$, and for each $k\in\mathcal{K}'$, we have
$$
V(k)=\frac{|\{x\in \mathbb{F}_2^n|F_k(x\oplus a)+F_k(x)=b\}|}{2^n}>1-q(n)t\epsilon.
$$
Let $\epsilon=\frac{\sigma}{q(n)t}$. Noticing that $p(n)=\frac{1}{2\sigma^2}q(n)^2n^3$, the probability of Eq.$(6)$ holding is greater than $1-ne^{-n}$. Therefore, except for a negligible probability, there exist a subset $\mathcal{K}'\subseteq\mathcal{K}$ satisfying $|\mathcal{K}'|/|\mathcal{K}|>1-\frac{1}{q(n)}$, and for any $k\in\mathcal{K}'$,
\begin{align*}
\frac{|\{x\in \mathbb{F}_2^n|F_k(x\oplus a)+F_k(x)=b\}|}{2^n}>1-q(n)t\epsilon=1-\sigma,
\end{align*}
which completes the proof.

$\hfill{} \Box$


When executing truncated differential cryptanalysis, the attacker first runs Algorithm 2. Except for a negligible probability, the output $(a,b)$ is a truncated differential of $F_k$ whose probability is greater than $1-\sigma$ for more than $(1-\frac{1}{q(n)})$ of the keys in $\mathcal{K}$. Then the attacker uses it to recover the subkey of the last round as in classical truncated differential cryptanalysis. This attack works for more than $(1-\frac{1}{q(n)})$ of the keys in $\mathcal{K}$. The number of pairs needed by the counting scheme is related to the signal to noise ratio $S/N$. The higher $S/N$ is, the fewer pairs of plaintexts are needed. It is observed experimentally that while $S/N$ is $1-2$, about 20-40 occurrences of right pairs are sufficient \cite{BS91}. (See Appendix A for more details.) Thus, about $\frac{40}{1-\sigma}$ pairs of plaintexts are enough.

\subsubsection{Complexity.} We analyze the complexity of Algorithm 2 from three perspectives: the number of universal gates, the time complexity of classical computing part and the amount of qubits needed.

\textbf{Amount of universal gates.} For each $j\in\{1,2,\cdots,n\}$, Algorithm 2 needs to execute BV algorithm on $F_j$ for $p(n)$ times, and each time needs $2(n+m)+1+|F_j|_Q$ universal gates. Thus, the total amount of needed universal gates is
\begin{align*}
&p(n)\sum_{j=1}^n\big[\,2(n+m)+1+|F_j|_Q\big]\\
=&p(n)\big[\,2n^2+(2m+1)n+\sum_{j=1}^n|F_j|_Q\big]\\
=&\frac{1}{2\sigma^2}q(n)^2n^3\big[\,2n^2+(2m+1)n+|F|_Q\big],
\end{align*}
which is a polynomial of $n$.

\textbf{Time complexity of classical computing part.} The classical computing part includes two phases:

(I) solving linear systems $\{x\cdot \omega=i_j|\omega\in H\}$ for all $j=1,2,\cdots,n$ and $i_j=0,1$;

(II) finding the intersection of the sets $A_j\,'s$.

For the first phase, the attacker needs to solve $2n$ linear systems, each one has $p(n)$ linear equations and $n$ variables. The complexity of solving a linear system with $\mu$ equations and $\nu$ variables by Gaussian elimination method is $O(\mu\nu^2)$. Thus, the complexity of the first phase is $O(2p(n)n^3)=O(\frac{1}{\sigma^2}q(n)^2n^6)$.

For the second phase, the attacker first lets $t=n$, if $A_1\cap\cdots\cap A_n\supsetneq\{\bm{0}\}$, then chooses an arbitrary nonzero vector $a$ in the intersection. Otherwise, the attacker lets $t=n-1$. If there exist $n-1$ sets $A_{j_1},\cdots,A_{j_{n-1}}$ such that $A_{j_1}\cap\cdots\cap A_{j_{n-1}}\supsetneq\{\bm{0}\}$, then he chooses an arbitrary nonzero vector $a$ in the intersection. Otherwise, he lets $t=n-2$. The attacker continues this process until the value of $t$ is too small to satisfy the condition that $S/N=2^t(1-\sigma)>1$, or finding the intersection of $t$ sets requires too many calculations. Supposing $\alpha=\max_j|A_j|$, finding the intersection of $t$ sets by sort method needs $O(t\alpha\log\alpha)$ calculations. Thus, selecting $t$ sets from $A_j\,'s$ that have nonzero common vectors needs $O(\binom{n}{t}t\alpha\log\alpha)$ calculations. Let $t_0$ be the minimum $t$ that satisfies $2^t(1-\sigma)>1$, then the complexity of the second phase is $O(\sum_{t=t_0}^n\binom{n}{t}t\alpha\log{\alpha})$. The value of $\alpha$ is determined by the properties of the attacked block cipher. It is generally small because a well constructed cipher should not have many approximate linear structures. Furthermore, the attacker can reduce the value of $\alpha$ by choosing a larger $p(n)$. In practice, the attacker can choose a polynomial $g(n)$ to represent the upper bound of his computational power. As long as $\binom{n}{t}t\alpha\log{\alpha}$ is greater than $g(n)$, he stops the process. In this situation the complexity of the second part is $O(ng(n))$.

To sum up, the complexity of classical computing part is $O(\frac{1}{\sigma^2}q(n)^2n^6+\sum_{t=t_0}^n\binom{n}{t}t\alpha\log{\alpha})$. If the attacker chooses a polynomial $g(n)$ to bound the amount of calculations, and tries from $t=n,n-1,n-2\cdots$ until $2^t(1-\sigma)\leq1$ or $\binom{n}{t}t\alpha\log{\alpha}$ is larger than $g(n)$, then the complexity of classical computing part is $O(\frac{1}{\sigma^2}q(n)^2n^6+ng(n))$, which is a polynomial of $n$.

\textbf{Amount of qubits needed.} Running BV algorithm on each $F_j$ needs $n+m+1$ qubits. Since the executions of BV algorithm is sequential, these qubits can be reused. Therefore, there are totally $n+m+1$ qubits needed to perform Algorithm 2.
\subsubsection{Advantages}

To illustrate the advantages of Algorithm 2, we compare it with the traditional method of finding high-probability truncated differentials. In conventional differential cryptanalysis, the attacker finds high-probability differentials by searching for high-probability differential characteristics. A differential characteristic is a sequence of input and output differences of the rounds satisfying that the input difference of one round equals to the output difference from the last round. In order to find high-probability differential characteristics, the attacker examines the properties of each S-boxes and looks for their high-probability differentials individually. Combining the difference pairs of S-boxes from round to round, the attack can find a high-probability differential characteristic containing the plaintext difference and the difference into the last round. The way to find high-probability truncated differentials in classical truncated differential cryptanalysis is similar, except that the input difference and output difference of each round can be only partially determined. Under the assumption that the differentials of different S-boxes are independent (which does not hold strictly, but works well for most block ciphers in practice), the probability of a truncated differential characteristic is equal to the product of the differential probabilities of all active S boxes. As the number of rounds increases, the number of active S-boxes will also increases, so the probability of the differential characteristic will be greatly reduced. Therefore, finding high-probability truncated differential characteristics usually becomes more and more difficult as the number of rounds increases. By contrast, Algorithm 2 only concerns the input and output differences at both ends of the reduced cipher $F_k$, instead of a specific differential characteristic, so the increase in the number of rounds has a much smaller effect on Algorithm 2. Therefore, compared with traditional truncated differential cryptanalysis, Algorithm 2 is more conductive to finding high-probability truncated differentials when the number of the rounds of the block cipher is large.

Another advantage of Algorithm 2 is that it can find more general high-probability truncated differentials than conventional methods. As we know, a high-probability truncated differential does not imply a high-probability truncated differential characteristic, because there may be multiple differential characteristics matching this truncated differential, but the probability of each one is not high. The high-probability truncated differentials whose input and output differences can be connected by a high-probability characteristic is actually restrictive. As analyzed earlier, the traditional method of finding high-probability truncated differentials is based on finding high-probability truncated differential characteristics, so the truncated differentials it can find are restrictive. By contrast, Algorithm 2 only concerns the input and output differences at both ends, and thus can find more general high-probability truncated differentials.  

No need for quantum queries is also an advantage of Algorithm 2. Being able to query the cryptographic oracle in quantum superpositions is a strong requirement for the attacker's ability. Many recently proposed quantum algorithms for attacking symmetric systems require the ability of querying with superpositions \cite{KLLNP16,KM12,RS15}. Compared with these algorithms, Algorithm 2 has less demanding requirement for attacker's ability, and therefore is easier to implement in practice.

\section{Quantum Impossible differential cryptanalysis}

Impossible differential cryptanalysis is a chosen-plaintext attack introduced by Biham, Biryukov, and Shamir \cite{BBS99}. While ordinary differential cryptanalysis makes use of high-probability differentials, impossible differential cryptanalysis exploits the differentials of probability zero.

We still consider a $r$-round block cipher $E$ and its reduced version $F=E^{(r-1)}$. Let $\mathcal{K}$ and $\mathcal{S}$ be the key spaces of the first $r-1$ rounds and the last round, respectively. When fix a specific key $k\in\mathcal{K}$, the action of $F$ on a block $x$ is denoted by $F_k(x)$. If a differential $(\Delta x,\Delta y)$ of $F_k$ satisfies that
$$
F_k(x\oplus \Delta x)\oplus F_k(x)\neq \Delta y, \,\,\forall x\in \mathbb{F}_2^n,
$$
then it is called an impossible differential of $F_k$. Impossible differential cryptanalysis consists of two phases: (I) finding some impossible differential $(\Delta x,\Delta y)$ of $F_k$; and (II) sieving the subkey of the last round based on the found impossible differential. In this paper, we focus on the first phase of impossible differential cryptanalysis, and propose a quantum algorithm for finding impossible differentials. The proposed algorithm applies the miss-in-the-middle technique. We present it in Section 4.1, and analyze its validity, complexity and advantages in Section 4.2.

\subsection{Quantum algorithm for finding impossible differentials}

The miss-in-the-middle technique \cite{BBS99} has been widely used in traditional impossible differential cryptanalysis. It has been applied to Skipjack \cite{BBS99}, IDEA \cite{BBS999}, DEAL \cite{Knu96} and so on. The basic idea of miss-in-the-middle technique is to connect two differential paths of probability one, whose corresponding input and output differences do not match, to obtain an impossible differential. Specifically, for $v\in\{1,\cdots,r-2\}$, we divide the reduced cipher $F_k$ into two parts: $F_k=\check{F}_{k_2}^{(v)}\cdot\hat{F}_{k_1}^{(v)}$, where $\hat{F}_{k_1}^{(v)}$ corresponds to the first $v$ rounds of $F_k$, $\breve{F}_{k_2}^{(v)}$ corresponds to the last $r-1-v$ rounds, and $k=(k_1,k_2)$. The key space $\mathcal{K}$ is accordingly divided into two parts $\mathcal{K}=\mathcal{K}_1^v\otimes\mathcal{K}_2^v$. If $(\Delta x_1,\Delta y_1)$ and $(\Delta x_2,\Delta y_2)$ are probability-1 differentials of $\hat{F}_{k_1}^{(v)}$ and $\check{F}_{k_2}^{(v)}$, respectively, and $\Delta y_1\neq\Delta x_2$, then $(\Delta x_1,\Delta y_2)$ will be an impossible differential of $F_k$. The miss-in-the-middle technique translates the task of finding impossible differentials into the task of finding probability-1 differentials.

In light of the fact that a linear structure of a Boolean function can induce a probability-1 differential of it, we can apply BV algorithm to find probability-1 differentials of  $\hat{F}_{k_1}^{(v)}$ and $\check{F}_{k_2}^{(v)}$. For example, suppose $\hat{F}_{k_1}^{(v)}=(\hat{F}_{k_1,1}^{(v)},\hat{F}_{k_1,2}^{(v)},\cdots,\hat{F}_{k_1,n}^{(v)})$. We can obtain a probability-1 differential of $\hat{F}_{k_1}^{(v)}$ by first using Algorithm 1 to find the linear structures of each $\hat{F}_{k_1,j}^{(v)}$ separately, and then choosing a common linear structure as the input difference. However, there is a problem that the attacker has no access to the quantum oracle of $\hat{F}^{(v)}_{k_1,j}$ due to his ignorance of $k_1$. To solve this problem, we employ the function $F$ without specifying the key as in the case of truncated differential cryptanalysis, and tries to find key-independent impossible differentials. Specifically, for $v\in\{1,2,\cdots,r-2\}$, we also divide $F$ into two parts:

\begin{align*}
\hat{F}^{(v)}:\{0,1\}^n\times\mathcal{K}_1^v&\rightarrow\{0,1\}^n\qquad&\check{F}^{(v)}:\{0,1\}^n\times\mathcal{K}_2^v\rightarrow\{0,1\}^n\\
(x,k_1)\,\,\,&\rightarrow \hat{F}_{k_1}^{(v)}(x),    &(x,k_2)\,\,\,\rightarrow \check{F}_{k_2}^{(v)}(x).
\end{align*}

Then we have $F(x,(k_1,k_2))=\check{F}^{(v)}(\hat{F}^{(v)}(x,k_1),k_2)$. In the following, we let $m,l_v$, and $h_v$ denote the lengths of the keys in $\mathcal{K}$, $\mathcal{K}_1^v$ and $\mathcal{K}_2^v$, respectively, where $l_v+h_v=m$. Suppose $\hat{F}^{(v)}=(\hat{F}_1^{(v)},\cdots,\hat{F}_n^{(v)})$, $\check{F}^{(v)}=(\check{F}_1^{(v)},\cdots,\check{F}_n^{(v)})$. Each $\hat{F}_j^{(v)}$ or $\check{F}_j^{(v)}$ is deterministic and known to the attacker, so he can use efficient quantum circuits to implement them by himself. According to Theorem 2, by running Algorithm 1 on each $\hat{F}_j^{(v)}$, the attacker is expected to obtain linear structures of $\hat{F}_j^{(v)}$. Since the input of $\hat{F}_j^{(v)}$ includes a plaintext block and a key in $\mathcal{K}_1^v$, the linear structures he obtains have $n+l_v$ bits, including $n$-bit plaintext difference and $l_v$-bit key difference. In order to use it to induce a probability-1 differential of $\hat{F}_{k1}^{(v)}$, the last $l_v$ bits of the linear structures need to be zeros. To do this, we only need to discard the last $l_v$ bits of the output vectors when calling BV Algorithm, as done in Algorithm 2. Then the length of the vectors obtained by solving the linear systems will be $n$ bits, and thus can be used to induced probability-1 differentials of $\hat{F}_{k1}^{(v)}$. For $v=1,2,\cdots,r-2$, the attacker uses this method to search for probability-1 differentials of $\hat{F}_{k_1}^{(v)}$ and $\check{F}_{k_2}^{(v)}$ separately until two unmatched probability-1 differentials are found.

The quantum algorithm for finding impossible differentials of $F_k$ is as following:

\begin{breakablealgorithm}[H]
\caption{}
\begin{algorithmic}[1]
\Require Quantum circuits for implementing $\hat{F}_j^{(v)}$ and $\check{F}_j^{(v)}$ ($v=1,\cdots,r-2$, $j=1,\cdots,n$) are given. $p(n)$ is an arbitrary polynomial chosen by the attacker. All appearing sets are initialized to the null set $\Phi$.
\For{$v=1,2,\cdots,r-2$}
 \For{$j=1,2,\cdots,n$}
   \For {$p=1,2,\cdots,p(n)$}
   \State run BV algorithm on $\hat{F}_j^{(v)}$ to obtain an $(n+l_v)$-bit output $\omega=$
   \State $(\omega_1,\cdots,\omega_n,\omega_{n+1},\cdots,\omega_{n+l_v})\in N_{\hat{F}_j^{(v)}}$;
   \State let $H=H\cup\{(\omega_1,\cdots,\omega_n)\}$;
   \EndFor
 \State solve the system of linear equations $\{x\cdot\omega=t_{v,j}|\omega\in H\}$ to obtain the
 \State sets $A_{v,j}^{t_{v,j}}$ for $t_{v,j}=0,1$, respectively; Let $A_{v,j}=A_{v,j}^0\cup A_{v,j}^1$;
 \If{ $A_{v,j}\subseteq\{\bm{0}\}$} \State\textbf{break}; (Exit current loop)
 \Else\State let $H=\Phi$;
 \EndIf
 \EndFor
 \If{ $A_{v,1}\cap\cdots\cap A_{v,n}\subseteq\{\bm{0}\}$}  \State\textbf{continue}; (Jump to the next iteration of current loop)
 \Else \State choose an arbitrary nonzero vector $a\in A_{v,1}\cap\cdots\cap A_{v,n}$;
       \State let $(\Delta x_1,\Delta y_1)=(a,t_{v,1},\cdots,t_{v,n})$, where $t_{v,1},\cdots,t_{v,n}$ are the super-
       \State scripts such that $a\in A_{v,1}^{t_{v,1}}\cap A_{v,2}^{t_{v,2}}\cap\cdots\cap A_{v,n}^{t_{v,n}}$;
 \EndIf

 \For{$j=1,2,\cdots,n$}
   \For{ $p=1,2,\cdots,p(n)$}
     \State run BV algorithm on $\check{F}_j^{(v)}$ to obtain an $(n+h_v)$-bit output $\omega=$
     \State $(\omega_1,\cdots,\omega_n,\omega_{n+1},\cdots,\omega_{n+h_v})\in N_{\check{F}_j^{(v)}}$;
     \State let $H=H\cup\{(\omega_1,\cdots,\omega_n)\}$;
   \EndFor
   \State solve the system of linear equations $\{x\cdot \omega=s_{v,j}|\omega\in H\}$ to obtain the
   \State sets $B_{v,j}^{s_{v,j}}$ for $s_{v,j}=0,1$, respectively; Let $B_{v,j}=B_{v,j}^0\cup B_{v,j}^1$;
   \State Let $H=\Phi$;
 \EndFor
 \If{$B_{v,1}\cap\cdots\cap B_{v,n}=\Phi$}
  \State\textbf{continue};
 \Else \State choose an arbitrary vector $b\in B_{v,1}\cap\cdots\cap B_{v,n}$;
 \State let $(\Delta x_2,\Delta y_2)=(b,s_{v,1},\cdots,s_{v,n})$, where $s_{v,1},\cdots,s_{v,n}$ are the super-
 \State scripts such that $b\in B_{v,1}^{s_{v,1}}\cap B_{v,2}^{s_{v,2}}\cap\cdots\cap B_{v,n}^{s_{v,n}}$;
 \EndIf
 \If{$\Delta y_1\neq\Delta x_2$} \State output $(\Delta x_1,\Delta y_2,0)$ and halt;
 \Else \State output $(\Delta x_1,\Delta y_2,1)$ and halt;
 \EndIf
\EndFor
 \end{algorithmic}
\end{breakablealgorithm}

Lines 2-22 of Algorithm 3 are for finding a probability-1 differential $(\Delta x_1,\Delta y_1)$ of $\hat{F}_{k_1}^{(v)}$, while lines 23-39 are for finding a probability-1 differential $(\Delta x_2,\Delta y_2)$ of $\check{F}_{k_2}^{(v)}$. Once probability-1 differentials of these two functions are found simultaneously for some $v$, the algorithm halts. When the attacker executes an impossible differential attack, he first runs Algorithm 3 with $p(n)=n$. According to Theorem 4 presented in the next subsection, as long as the block cipher satisfies certain properties, the vector output by Algorithm 3 will be either an impossible differential or a probability-1 differential of $F_k$. Specifically, if the attacker obtains an output $(\Delta x_1,\Delta y_2,0)$, then except for a negligible probability, $(\Delta x_1,\Delta y_2)$ is an impossible differential of $F_k$. The attacker can use it to sieve the key of the last round as in classical impossible differential cryptanalysis. If he obtains $(\Delta x_1,\Delta y_2,1)$, then except for a negligible probability, $(\Delta x_1,\Delta y_2)$ is a probability-1 differential of $F_k$. This situation is actually more conducive to recover the key of the last round, but the probability of it happening is usually very small because a well constructed cipher usually does not have such strong linearity. Unlike the case of truncated differential cryptanalysis, no matter the vector obtained via Algorithm 3 is an impossible differential or a probability-1 differential of $F_k$, its probability is independent of the key.

\subsection{Analysis of the algorithm}
In this section, we discuss the validity, complexity and advantages of Algorithm 3.

\subsubsection{Validity} To demonstrate the validity of Algorithm 3, we first define the parameter
$$
\delta'_F=\max\{\delta'_{\hat{F}_j^{(v)}},\delta'_{\check{F}_j^{(v)}}|1\leq v\leq r-2,1\leq j\leq n \},
$$
where $\delta'_{\hat{F}_j^{(v)}},\delta'_{\check{F}_j^{(v)}}$ is defined as in Eq.$(2)$. It is obvious that $\delta'_F<1$. The smaller $\delta'_F$ is, the better it is to rule out the vectors that are not the linear structure of $\hat{F}^{(v)}$ or $\check{F}^{(v)}$ when running Algorithm 3. The following theorem justifies the validity of Algorithm 3:
\begin{theorem}Suppose $\delta'_F\leq p_0<1$ for some constant $p_0$. If applying Algorithm 3 on $F_k$ with $p(n)=n$ outputs $(\Delta x_1,\Delta y_2,0)$, then except for a negligible probability, $(\Delta x_1,\Delta y_2)$ is an impossible differential of $F_k$ for all $k\in\mathcal{K}$. If it outputs $(\Delta x_1,\Delta y_2,1)$, then except for a negligible probability, $(\Delta x_1,\Delta y_2)$ is a probability-1 differential of $F_k$ for all $k\in\mathcal{K}$.
\end{theorem}

\noindent
\textbf{Proof}. Since $a\cdot(\omega_1,\cdots,\omega_n)=0$, the vector $(a\|0,\cdots,0)$, obtained by appending $l_v$ zeros to $a$, satisfies that $(a\|0,\cdots,0)\cdot(\omega_1,\cdots,\omega_n,\omega_{n+1},\cdots,\omega_{n+l_v})=0$. Thus,  $(a\|0,\cdots,0)$ can be viewed as the output when applying Algorithm 1 to $\hat{F}_j^{(v)}$. Then according to Theorem 2, for all $1\leq v\leq r-2$, all $1\leq j\leq n$, it holds that
$$
{\rm Pr}[\,(a\|0,\cdots,0)\notin U_{\hat{F}_j^{(v)}}^{t_{v,j}}]\leq p_0^n.
$$
Since $(a\|0,\cdots,0)\notin U_{\hat{F}^{(v)}}^{(t_{v,1},\cdots,t_{v,n})}$ implies that there exists some $j_0$ such that $(a\|0,\cdots,0)\notin U_{\hat{F}_{j_0}^{(v)}}^{t_{v,j_0}}$, we have
$$
{\rm Pr}[\,(a\|0,\cdots,0)\notin U_{\hat{F}^{(v)}}^{(t_{v,1},\cdots,t_{v,n})}]\leq p_0^n.
$$
When $(a\|0,\cdots,0)\in U_{\hat{F}^{(v)}}^{(t_{v,1},\cdots,t_{v,n})}$, we have that for all $x\in \mathbb{F}_2^n$ and all $k_1\in\mathcal{K}_1^v$,
$$
\hat{F}^{(v)}\Big((x,k_1)\oplus(a\|0,\cdots,0)\Big)\oplus\hat{F}^{(v)}\big(x,k_1\big)=(t_{v,1},\cdots,t_{v,n}).
$$
That is,
$$
\hat{F}^{(v)}_{k_1}(x\oplus a)\oplus\hat{F}^{(v)}_{k_1}(x)=(t_{v,1},\cdots,t_{v,n}),\,\,\forall x\in \mathbb{F}_2^n, \forall k_1\in\mathcal{K}_1.
$$
Thus, except for a negligible probability, $(a,t_{v,1},\cdots,t_{v,n})$ is a probability-1 differential of $\hat{F}_{k_1}^{(v)}$ for all $k_1\in\mathcal{K}_1^v$. Similarly, except for a negligible probability, $(b,s_{v,1},\cdots,s_{v,n})$ is a probability-1 differential of $\check{F}_{k_2}^{(v)}$ for all $k_2\in\mathcal{K}_2$. Since $F_k=\check{F}_{k_2}^{(v)}\cdot\hat{F}_{k_1}^{(v)}$, the conclusion holds.

$\hfill{} \Box$

According to Theorem 4, as long as $\delta'_F\leq p_0<1$ for some constant $p_0$ and running Algorithm 3 on $F$ returns a vector, then the vector will be an impossible differential or a probability-1 differential of $F_k$ except for a negligible probability. But Theorem 4 does not give the condition under which Algorithm 3 must output a vector. In fact, under the assumption that $\delta'_F\leq p_0<1$, we have stronger conclusion: as long as $F_k$ has an impossible differential that is composed of two unmatched probability-1 differentials, Algorithm 3 will, except for a negligible probability, output an impossible differential or a probability-1 differential of $F_k$.

To explain why this holds, we suppose $F_k$ has an impossible $(\Delta x, \Delta y)$ that is composed of two unmatched probability-1 differentials. Then there must exist $v\in\{1,\cdots,r-2\}$, $\Delta y_1\in\mathbb{F}_2^n$ and $\Delta x_2\in\mathbb{F}_2^n$ such that $(\Delta x,\Delta y_1)$ and $(\Delta x_2,\Delta y)$ are probability-1 differentials of $\hat{F}_{k_1}^{(v)}$ and $\check{F}^{(v)}_{k_2}$ respectively, and $\Delta y_1\neq \Delta x_2$. In this case, according to Lemma 1, $\Delta x$ must be a solution of the linear systems $\{x\cdot\omega=\Delta y_{1,j}|\omega\in H\}$ for all $j=1,\cdots,n$, where $\Delta y_{1,j}$ is the $j^{th}$ bit of $\Delta y_{1}$. Thus, $\Delta x$ must be in the set $A_v\triangleq A_{v,1}\cap\cdots\cap A_{v,n}$. Likewise, $\Delta x_2$ must be in the set $B_v\triangleq B_{v,1}\cap\cdots\cap B_{v,n}$. Therefore, as long as $F_k$ has an impossible differential that is composed of two unmatched probability-1 differentials, there must exist some $v$ such that the input differences of these two probability-1 differentials are in the sets $A_v$ and $B_v$, respectively. On the other hand, according to Theorem 4, as long as the reduced cipher $F$ satisfies that $\delta'_F\leq p_0<1$ for some constant $p_0$, the probability of the vectors in $A_v$ (resp. $B_v$) are not linear structures of $\hat{F}_{k_1}^{(v)}$ (resp. $\check{F}^{(v)}_{k_2}$) is negligible. Therefore, as long as $A_v$ and $B_v$ are both nonempty, the vectors chosen respectively from them will form an impossible differential or a probability-1 differential of $F_k$ except for a negligible probability. This justifies the above conclusion.

Based on the above analysis, if the classical miss-in-the-middle technique works for some reduced cipher $F_k$, which means $F_k$ must have an impossible differential that is composed of two unmatched probability-1 differentials, then except for a negligible probability, Algorithm 3 will find out an impossible differential or a probability-1 differential of $F_k$. Therefore, to a certain extent, we can say that as long as classical miss-in-the-middle technique works for some block cipher $E$, which satisfies $\delta'_{E^{(r-1)}}\leq p_0<1$ for some constant $p_0$, then Algorithm 3 must work for it too.

\subsubsection{Complexity}  We analyze the complexity of Algorithm 3 from three perspectives: the number of universal gates, the time complexity of classical computing part and the amount of qubits needed.

\textbf{Amount of universal gates.} For each $v=1,2,\cdots,r-2$ and $j=1,2,\cdots,n$, Algorithm 3 needs to execute BV algorithm on $\hat{F}_j^{(v)}$ and $\check{F}^{(v)}_j$ for $p(n)$ times, and each time needs $2(n+l_v)+1+|\hat{F}^{(v)}_j|_Q$ and $2(n+h_v)+1+|\check{F}_j^{(v)}|_Q$ universal gates, respectively. Thus, the total amount of needed universal gates is
\begin{align*}
&p(n)\sum_{v=1}^{r-2}\sum_{j=1}^n\Big[2(n+l_v)+1+|\hat{F}^{(v)}_j|_Q+2(n+h_v)+1+|\check{F}_j^{(v)}|_Q\Big]\\
=&p(n)\sum_{v=1}^{r-2}\sum_{j=1}^n\Big[4n+2+2(l_v+h_v)+|\hat{F}^{(v)}_j|_Q+|\check{F}_j^{(v)}|_Q\Big]\\
=&p(n)\sum_{v=1}^{r-2}\sum_{j=1}^n(4n+2+2m)+p(n)\sum_{v=1}^{r-2}\Big(\sum_{j=1}^n|\hat{F}_j^{(v)}|_Q+\sum_{j=1}^n|\check{F}_j^{(v)}|_Q\Big)\\
=&p(n)(r-2)n(4n+2+2m)+p(n)\sum_{v=1}^{r-2}\big(|\hat{F}^{(v)}|_Q+|\check{F}^{(v)}|_Q\big)\vspace{2.8ex}\\
=&p(n)(r-2)n(4n+2+2m)+p(n)\sum_{v=1}^{r-2}|F|_Q\\[0.1cm]
=&p(n)(r-2)(4n^2+2n+2mn+|F|_Q)\\[0.3cm]
=&(r-2)(4n^3+2n^2+2mn^2+n|F|_Q),
\end{align*}
which is a polynomial of $n$. The last formula holds since $p(n)=n$.
\vskip 0.07cm

\textbf{Time complexity of classical computing part.} The classical computing part of Algorithm 3 includes two phases:

\hangafter 1
\hangindent 1.8em
(1) Solve linear systems $\{x\cdot\omega=t_{v,j}|\omega\in H\}$ and $\{x\cdot\omega=s_{v,j}|\omega\in H\}$ for each $v\in\{1,2\cdots,r-2\}$, $j\in\{1,2,\cdots,n\}$ and $t_{v,j}, s_{v,j}\in\{0,1\}$;

(2) Find the intersection of the sets $A_{v,j}\,'s$ and $B_{v,j}\,'s$.

For the first phase, the attacker needs to solve $4(r-2)n$ linear systems, each one has $p(n)$ linear equations and $n$ variables. Thus, the complexity of this phase is $O(4(r-2)np(n)n^2)=O(4(r-2)n^4)$. For the second phase, the corresponding complexity is determined by the size of the sets $A_{v,j}\,'s$ and $B_{v,j}\,'s$. Let $\alpha=\max_{v,j}\{|A_{v,j}|,|B_{v,j}|\}$. Since the attacker needs to take intersection for at most $2(r-2)$ times, the complexity of this phase is $O(2(r-2)n\alpha\log \alpha)$. Therefore, the total time complexity of classical computing part is $O\big((r-2)(2n^4+n\alpha\log\alpha)\big)$. In general, the value of $\alpha$ is small because a well constructed cipher should not have many approximate linear structures. Furthermore, the attacker can reduce the value of $\alpha$ by choosing a larger $p(n)$.

\textbf{Amount of qubits needed.} Running BV algorithm on each $\hat{F}_j^{(v)}$ or $\check{F}_j^{(v)}$ needs $n+l_v+1$ or $n+h_v+1$ qubits, respectively. Since both $l_v$ and $h_v$ are not larger than $m$ and these qubits can be reused, $n+m+1$ qubits are enough to perform Algorithm 3.

\subsubsection{Advantages} To illustrate the advantages of Algorithm 3, we compare it with classical miss-in-the-middle technique. The basic idea of miss-in-the-middle technique is to find impossible differentials by searching for two unmatched probability-1 differentials. In classical case, the attacker finds probability-1 differentials by looking for probability-1 differential characteristics. As we analyzed in the case of truncated differential cryptanalysis, the probability of differential characteristics will decrease greatly as the number of rounds increases. Thus, finding probability-1 differential characteristics will become more and more difficult as the number of rounds increases. By contrast, Algorithm 3 treats $\hat{F}^{(v)}_j$ and $\check{F}^{(v)}_j$ as a whole and only cares the input and output differences at both ends of them. Whether Algorithm 3 works or not does not depend on the existence of probability-1 differential characteristics. Therefore, compared with the traditional miss-in-the-middle technique, the increase in the number of rounds has a much smaller effect on Algorithm 3.

In addition, the existence of a probability-1 differential does not imply the existence of a probability-1 differential characteristic. The probability-1 differentials whose input and output differences can be connected by a probability-1 differential characteristic are actually restrictive. As discussed above, the impossible differentials found by classical miss-in-the-middle technique are always connected by two probability-1 differential characteristics, while the impossible differentials found by Algorithm 3 are connected by two general probability-1 differentials, without other restrictions. Therefore, to some extent we can say that the impossible differentials that Algorithm 3 can find are more general than the impossible differentials that traditional miss-in-the-middle technique can find. Furthermore, as analyzed in Subsection 4.2.1, for any block cipher $E$ that satisfies $\delta'_{E^{(r-1)}}\leq p_0<1$ for some constant $p_0$, if classical miss-in-the-middle technique can find an impossible differential of it, Algorithm 3 must be able to find an impossible (or a probability-1 differential) of it, too.

No need for quantum queries is also an advantage of Algorithm 3. This reduces the requirement for the attacker's ability, and makes the attack on block ciphers more practical. Compared with Algorithm 2, Algorithm 3 also has the advantage that the impossible differentials found by it are key-independent, while the truncated differentials found by Algorithm 2 have high probability only for partial keys.

\section{Conclusions and further directions}

In this paper, we apply Bernstein-Vazirani algorithm in truncated differential and impossible differential cryptanalysis. We propose two quantum algorithms that can be used to find high-probability truncated differentials and impossible differentials, respectively. We believe our work contributes to a better understanding of the impact of quantum computing on symmetric cryptanalysis, and provides guidance for the design of quantum-secure symmetric cryptosystems.

There are still many directions to further investigate. First, it may be possible to improve the proposed quantum algorithm for finding truncated differentials so that it can find key-independent high-probability truncated differentials. Also, how to take the key-recovery process into account for finding the optimized high-probability truncated differentials or impossible differentials is worth further studying. In addition, applying Bernstein-Vazirani algorithm to other variants of differential cryptanalysis, such as higher-order differential cryptanalysis and boomerang attacks, may leads to interesting results.

\section*{Acknowledgement}
This work was supported by National Natural Science Foundation of China (Grant No.61672517), National Cryptography Development Fund (Grant No. MMJJ201 70108) and the Fundamental theory and cutting edge technology Research Program of Institute of Information Engineering, CAS (Grant No. Y7Z0301103).

\begin{appendix}
\section{Signal to noise ratio}

In this section we briefly recall the notion of the signal to noise ratio \cite{BS91}, which gives us a tool for evaluating the usability of a counting scheme based on a high-probability differential.

In the counting scheme of a (truncated) differential cryptanalysis, the attacker uses a given high-probability (truncated) differential $(\Delta x,\Delta y)$ of the reduced cipher $F_k$ to recover the key of the last round. Specifically, let $\mathcal{S}$ be the key space of the last round of the block cipher. The attacker first fixes the input difference $\Delta x$ and makes $2N$ classical queries to obtain $2N$ ciphers. Then for each $s\in\mathcal{S}$, he decrypts the last round to get $N$ output differences of $F_k$, and counts the number of them that match $\Delta y$. The candidate key with the maximum count is chosen as the key of the last round. The definition of the signal to noise ratio is as follows:
\begin{definition}[\cite{BS91}, Definition 13] The \textit{signal to noise ratio of the counting scheme}, denoted by $S/N$, is the ratio between the number of times the right key is counted and the average times a random key is counted.
\end{definition}

Let $\gamma$ be the average count contributed by each plaintext pair and $\lambda$ be the ratio of non-discarded pairs to all pairs. (There may be a procedure to discarded the wrong pairs before they are actually counted.) Then the average times a random key is counted is $N\cdot\lambda\cdot\gamma/|\mathcal{S}|$, where $N$ is the number of pairs. Supposing that $p$ is the probability of the used differential (or truncated differential), then the number of times the right key is counted is about $N\cdot p$. Therefore, the signal to noise ratio is
$$
S/N=\frac{N\cdot p}{N\cdot\lambda\cdot\gamma/|\mathcal{S}|}=\frac{|\mathcal{S}|\cdot p}{\lambda\cdot\gamma}.
$$
If $S/N\leq1$, then the differential attack will not succeed.

In the counting scheme, if a pair of ciphertexts, after being decrypted by the correct key of the last round, matches the output difference of the given high-probability (truncated) differential, then this pair of ciphertexts is called the right pair. The amount of pairs required by a counting scheme is usually related to the amount of right pairs required, which is basically a function of the signal to noise ratio. If $S/N$ is sufficiently large, only a few occurrences of right pairs are enough to determine the value of the key of the last round. It is observed experimentally that, while $S/N$ is $1-2$, about $20-40$ right pairs are enough \cite{BS91}. While $S/N$ is much higher, only $3-4$ right pairs are sufficient. On average, about $O(\frac{1}{p})$ pairs of ciphertexts gives a right pair. Thus, when $S/N$ is $1-2$, the number of pairs needed is $O(\frac{40}{p})$. To further understand the relation between $S/N$ and the complexity of the counting scheme, readers are referred to \cite{BS91,Knu94} for concrete examples.
\end{appendix}

\begin{thebibliography}{}

\bibitem{BV97} Bernstein, E., Vazirani, U.: Quantum complexity theory. SIAM Journal on Computing, 26(5), 1411¨C1473 (1997)

\bibitem{BBS99} Biham, E., Biryukov, A., $\&$ Shamir, A.: Cryptanalysis of Skipjack reduced to 31 rounds using impossible differentials. In: International Conference on the Theory and Applications of Cryptographic Techniques, pp. 12¨C23 (1999)

\bibitem{BBS999} Biham, E., Biryukov, A., $\&$ Shamir, A.: Miss in the middle attacks on IDEA and Khufu. In: Fast Software Encryption-FSE'99, pp.124¨C138 (1999)

\bibitem{BS91}Biham, E., Shamir, A.: Differential cryptanalysis of {DES}-like cryptosystems. Journal of CRYPTOLOGY, 4(1), 3¨C72 (1991)

\bibitem{BZ13}Boneh D., Zhandry M.: Secure signatures and chosen ciphertext security in a quantum computing world. In: CRYPTO 2013, Part II, pp. 361¨C379 (2013)

\bibitem{Dub01} Dubuc, S.: Characterization of linear structures. Designs, Codes and Cryptography, 22(1), 33¨C45 (2001)

\bibitem{GHS16}Gagliardoni T., Hlsing A., Schaffner C.: Semantic security and indistinguishability in the quantum world. In: CRYPTO 2016, Part III, pp. 60¨C89 (2016)

\bibitem{Gro96} Grover, L. K.: A fast quantum mechanical algorithm for database search. In: Proceedings of the twenty-eighth annual ACM symposium on Theory of computing, pp. 212¨C219 (1996)


\bibitem{KLLNP16} Kaplan, M., Leurent, G., Leverrier, A., $\&$ Naya-Plasencia, M.: Breaking symmetric cryptosystems using quantum period finding. Advances in Cryptology¨CCRYPTO 2016, pp. 207¨C237 (2016)
\bibitem{KLLNP17} Kaplan, M., Leurent, G., Leverrier, A., $\&$ Naya-Plasencia, M.:  Quantum differential and linear cryptanalysis. In: Fast Software Encryption-FSE 2017 (2017)
\bibitem{Knu96}Knudsen, L. R.: DEAL-a 128-bit block cipher. complexity, 258(2), pp.216 (1998)
\bibitem{KB96} Knudsen, L. R., Berson, T. A.: Truncated differentials of SAFER. In: International Workshop on Fast Software Encryption, pp. 15¨C26 (1996)

\bibitem{KRW99} Knudsen, L. R., Robshaw, M. J.: Truncated differentials and Skipjack. Advances in Cryptology¨CCRYPTO 1999, pp. 165¨C180 (1999)


\bibitem{Knu94} Knudsen, L. R.: Truncated and higher order differentials. International. In: Workshop on Fast Software Encryption, pp.196¨C211 (1994)

\bibitem{KM10} Kuwakado, H., Morii, M.: Quantum distinguisher between the 3-round feistel cipher and the random permutation. Information Theory Proceedings (ISIT), 2010 IEEE International Symposium on, pp. 2682¨C2685 (2010)

\bibitem{KM12} Kuwakado, H., Morii, M.: Security on the quantum-type even-mansour cipher. pp. 312¨C316 (2012)

\bibitem{LY18} Li H., Yang L.: A quantum algorithm to approximate the linear structures of Boolean functions. Math. Struct. Comput. Sci. 28, pp.1¨C13 (2018)



\bibitem{LY15} Li H., Yang L.: Quantum differential cryptanalysis to the block ciphers. In: International Conference on Applications and Techniques in Information Security, pp. 44¨C51 (2015)
\bibitem{MC88} Luby, M., Rackoff, C.: How to construct pseudorandom permutations from pseudorandom functions. SIAM J. Comput. 17(2), 373¨C386 (1988)

\bibitem{NC00} Nielsen, M., Chuang, I.: Quantum Computation and Quantum Information. Cambridge University Press, Cambridge (2000)

\bibitem{OK94} O'connor, L., Klapper, A.: Algebraic nonlinearity and its applications to cryptography. Journal of Cryptology, 7(4), pp. 213¨C227 (1994)

\bibitem{RS15}Roetteler, M., Steinwandt, R.: A note on quantum related-key attacks. Information Processing Letters, 115(1), pp.40-44 (2015)

\bibitem{SS17} Santoli, T., Schaffner, C.: Using simon¡¯s algorithm to attack symmetric-key cryptographic primitives. Quantum Information $\&$ Computation, 17(1$\&$2), pp.65¨C78 (2017)

\bibitem{Sho94} Shor, P. W.: Algorithms for quantum computation: Discrete logarithms and factoring. pp. 124¨C134 (1994)

\bibitem{Sim97} Simon, D. R.: On the power of quantum computation. SIAM journal on computing, 26(5), pp.1474¨C1483 (1997)

\bibitem{XY17} Xie, H., Yang, L.: Using Bernstein-Vazirani algorithm to attack block ciphers. Designs, Codes and Cryptography (2018). doi:10.1007/s10623-018-0510-5.

\bibitem{ZLZS15} Zhou, Q., Lu, S., Zhang, Z., $\&$ Sun, J.: Quantum differential cryptanalysis. Quantum Information Processing, 14(6), pp.2101¨C2109 (2015)




\end{thebibliography}
\end{document}